\documentclass[12pt]{article}
\begin{document}
\begin{center}
{
\bfseries
ON THE RELATIONS BETWEEN TWO-PHOTON
AND LEPTONIC
WIDTHS OF LOW-LYING S-WAVE STATES OF CHARMONIUM
}
\vskip 5mm

S.B.Gerasimov $^{1}$ and M.Majewski$^{2}$
\vskip 5mm

{\small
(1) {\it Bogoliubov Laboratory of Theoretical Physics, JINR, Dubna} \\
%\vspace*{.3cm}
(2) {\it University of Lodz, Department of Theoretical Physics, Lodz, Poland}\\
%$\dag$ {\it E-mail: gerasb@thsun1.jinr.ru}
}
\end{center}

\vskip 5mm

\begin{center}
\begin{minipage}{150 mm}
\centerline {\bf Abstract}

The relation between the ratio $\Gamma_{ee}(\psi^{'})/\Gamma_{ee}(J/\psi)$
and $\Gamma_{2\gamma}(\eta_{c}^{'})/\Gamma_{2\gamma}(\eta_{c})$, expressed
in terms of the configuration mixing amplitudes induced by the
contact spin-spin interaction of quarks in the ground and radial
excitation states, is shown to give, after the inclusion
of the newly derived relativistic corrections, the radiative
$\eta_{c}^{'}$-~ and $\eta_c$-width ratio in fair accord with recent
experiments. The dynamical model is proposed to derive
the ratio of relative probability of the ground~($\eta_{c}(2980)$)- and
first radial excitation ($\eta^{'}_{c}(3640)$)-state formation in
$\gamma \gamma$-collisions followed by their decay into the~
$\bar{K}K\pi$ and $p \bar{p}$ channels.
\end{minipage}
\end{center}
\vskip 10mm

{\bf 1.}~~Two-photon decays of heavy quarkonia provide valuable information
on heavy quark dynamics and have been under consideration
in recent experimental \cite{pdg02,cleo00,cleo03,babar03,belle02,e835,e760,l3-99,delphi98}
and theoretical \cite{gupta96,munz96,chao97,cza01,fab02,efg03,kim04}~studies.
It seems reasonable to assume, at least as a first approximation,
that there is no mixing between light and heavy quark sectors,
and that one can consider the needed transition amplitudes separately
for mesons constructed of the u-,d-, s-, and heavy c- and b-quarks.

In this work, we concentrate on the charmed quark sector and
our main concern in this problem will be
the question of the degree of model (in)depen- \\
dence of the $S$-wave $Q\bar Q$
annihilation rates, or their ratios, with respect to the role
of short-range spin-dependent forces.
This is an important theoretical question because in many approximate
relativistic
approaches to description of the annihilation of bound antiquark-quark
$S$-wave states that are subjected to strong short-range interactions
one needs to introduce the cut-off procedures to get rid of
singular behaviour of matrix elements considered, the cut-off
parameters or the "smearing" procedures being introduced
basically on the phenomenological grounds.

{\bf 2.}~~We start with just postulating for a mass
operator of the heavy quark $(Q=c,b)$~$S$-wave systems the simplest
spin-dependent matrix
\begin{eqnarray}
\langle n^{'}|\hat{M}|n \rangle = M_{n}\delta_{nn^{'}} +
C\cdot \psi_{nS}(0)\psi_{n^{'}S}(0)\cdot
\langle \vec{\sigma}_{Q}\cdot \vec{\sigma}_{\bar Q}\rangle
\label{mmatr}
\end{eqnarray}
and write down simple perturbation theory
expressions to have an estimation for the mutual change
of the ground and first radial
excited states due to switching on the contact spin-spin potential
\begin{eqnarray}
\Psi_{n}(0)=\psi_{n}(0)+\sum_{m}^{'}|m\rangle\langle m|\hat V |n\rangle
/(E_n^{(0)}-E_m^{(0)}),
\label{pt}
\end{eqnarray}
where $|i\rangle \equiv \psi_{i}$, and $E_{i}^{(0)}$ are the wave
functions and energies of the unperturbed hamiltonian and $\hat{V}$ is the
perturbation potential that we identify with the contact spin-spin
interaction of quarks.

Applying Eq.(\ref{pt}) successively to $\psi_{nS}(J)$, where $n=1,2$-
are the quantum numbers of the S-wave radial excitations of the $\bar{c}c$-
 quarkonia with the spin $J=0,1$, using the specific ({\it i.e.},
 proportional to the $\delta$-function form of the perturbation potential)
 and keeping everywhere the terms of the first order in ${\cal O}(\hat{V})$
we get the relations
\begin{eqnarray}
R_{2S/1S}^{J}|_{hfs} &\simeq&
\frac{\Psi_{2S}^{J}(0)^2}{\Psi_{1S}^{J}(0)^2}|_{hfs} \nonumber \\
&\simeq& \frac{\psi_{2S}(0)^2}{\psi_{1S}(0)^2}
\cdot \frac{[(1+V_{11}^{J}(m_{2}^{0}-m_{1}^{0})^{-1}+
V_{33}^{J}(m_{2}^{0}-m_{3}^{0})^{-1}+\ldots)]^2}
{[(1+V_{22}^{J}(m_{1}^{0}-m_{2}^{0})^{-1}+
V_{33}^{J}(m_{1}^{0}-m_{3}^{0}){-1}+\ldots)]^2} \nonumber  \\
&\simeq& \frac{\psi_{2S}(0)^2}{\psi_{1S}(0)^2}\cdot
[1+2(s_{2}^{J}- s_{1}^{J}],
\label{fraction}
\end{eqnarray}
where $s_{n}^{J},~J=0,1$~is the sum of terms of the order
${\cal O}(\hat V)$ entering into the denominator and numerator of (3).
Using the evident relation $V^{J=0}=-3V^{J=1}$ we obtain
\begin{eqnarray}
R_{2S/1S}^{J=0}|_{hfs}+3\cdot R_{2S/1S}^{J=1}|_{hfs} &\simeq&
4\cdot \frac{\psi_{2S}(0)^2}{\psi_{1S}(0)^2},
\label{hfs}
\end{eqnarray}
We make an estimation for the ratio
$\Gamma (\eta_{c}^{'}\rightarrow \gamma \gamma )/
\Gamma (\eta_{c}\rightarrow \gamma \gamma )$
on the basis of  experimental data of leptonic charmonium decays,
approximate validity of the lowest order perturbation theory for
the color-hyperfine splitting interaction as well as
on our anew obtained form of the relativistic corrections to the considered
decays.

We remind first the known results \cite{bgk76,km83} for
the lowest order QCD corrections
\begin{equation}
\frac {m_{\eta_{c}}^{2}\Gamma (\eta_c \rightarrow \gamma \gamma )}
{m_{J/\psi} ^{2}\Gamma(J/\psi
\rightarrow e^+e^-)}=\frac {4}{3} (1+1.96 \frac {\alpha_s}{\pi})
\frac {{\vert \Psi _{\eta_c}(0) \vert}^2}{{\vert
 \Psi_{J/\psi}(0)\vert}^2}\; \; ,
 \end{equation}
where $\alpha_s$ should be evaluated at the charm scale \cite{fab02}.

In certain chosen ratios such as
\begin{eqnarray}
\frac{\Gamma_{\gamma\gamma}^{\eta_{c}^{'}}\Gamma_{ee}^{J/\psi}}
{\Gamma_{\gamma\gamma}^{\eta_{c}} \Gamma_{ee}^{\psi^{'}}}=
(\frac{m_{\eta_{c}} m_{\psi^{'}}}
{m_{\eta_{c}^{'}} m_{J/\psi}}
)^2\frac{R_{2S/1S}^{J=0}}{R_{2S/1S}^{J=1}}
\label{double}
\end{eqnarray}
the ratios of $\psi_{nS}(0)$'s are cancelled and also
the QCD radiative corrections are assumed to be mutually compensated
not only in the next-to-leading order (NLO), but also in the higher orders of the perturbation
theory, which, as seen from refs.\cite{cza01,ben98,cza98}, are not negligible.
We assume further on that each type of studied corrections can be approximately
represented in factorized form
\begin{eqnarray}
\Psi_{nS}^{J}(0) \equiv |\psi_{nS}(0)|^{2}(1+\delta^J(rad)+\delta_{nS}^{J}(rel)+
\delta_{nS}^{J}(hfs))\simeq \nonumber \\
\simeq |\psi_{nS}(0)|^{2}(1+\delta^{J}(rad))(1+\delta_{nS}^{J}(rel))
(1+\delta_{nS}^{J}(hfs)).
\end{eqnarray}
That means that the $1+\delta^{J}(rad)$ factor coincides in the lowest
order radiative correction with $1+(a^{J}/\pi)\alpha_{s}(m_{c})$, where
$(a^{J=1}=-5.34)$ and $(a^{J=0}=-3.38)$,
%%%%%%%%%%%%%%%%%%%%%%%%
and it is assumed to be cancelled, together with higher order terms,
in both the leptonic ($R_{2S/1S}^{J=1}$) and 2-photon ($R_{2S/1S}^{J=0}$)
"single ratios".

The notation $\delta_{nS}^{J}(hfs)$~refers to the correction
factor due to the spin-spin potential-induced factors in (\ref{hfs})
and this type of corrections is not cancelled in the double ratio
(\ref{double}) unlike the mentioned linear combination of the "single
ratios" (\ref{hfs}).

The relativistic correction factor $1+\delta_{nS}^{J}(rel)$~is defined in
the following manner.
The static approximation for both $\Gamma_{ee}$ and $\Gamma_{\gamma \gamma}$,
resulting in their proportionality to the respective wave function value
"at origin", $\psi_{nS}(0)$, follows from neglecting of the dependence
of the bound quark annihilation amplitudes on their internal motion momenta
\begin{eqnarray}
|\int d\vec{p}~A(c\bar{c}(\vec {p}) \to ee(\gamma \gamma))~\phi_{nS}(\vec{p})|^2
\sim |\psi_{nS}(0)|^2 |A_{thr}(c\bar{c} \to ee(\gamma \gamma)|^2.
\end{eqnarray}
%%%%%%%%%%%%%%%%%%%%%
We have found that the appearance of the $\psi_{nS}(0)$ together
with the relativistic correction factor $1+\delta_{nS}^{J}(rel)$
can be justified from more general expression that includes
the non-static annihilation amplitudes
depending on the internal quark momenta taken into account.
\begin{eqnarray}
2\sqrt{1+\delta_{nS}^{J=0}(rel)} \simeq \frac{m_{c}^2}{\epsilon_{nS}^{2}}
\int \frac{d^{3}p}{(2\pi)^3}\phi_{nS}(p)
\frac{\epsilon_{nS}}{p}\log(\frac{\epsilon_{nS}+ p}{\epsilon_{nS}- p})
/\int\frac{d^{3}p}{(2\pi)^3}\phi_{nS}(p), \\
\sqrt{1+\delta_{nS}^{J=1}(rel)} \simeq
\int \frac{d^{3}p}{(2\pi)^3}\phi_{nS}(p)(1-\frac{\epsilon_{nS}- m_c}
{3\epsilon_{nS}})/\int\frac{d^{3}p}{(2\pi)^3}\phi_{nS}(p),
\label{ave}
\end{eqnarray}
where $\phi_{nS}(p)$ is the $nS$-state wave function in the momentum representation,
and $\epsilon_{nS}$ is the quark energy in the $nS$-state.
The momentum-dependent factors of the $(c\bar c)_{nS} \to e^{+}e^{-}
(\gamma \gamma)$ amplitudes are given, {\it e.g.}, in \cite{efg03},
but instead of taking $\epsilon(p)=(m_{c}^{2}+p^2)^{1/2}$
we prefer to define the continuation of the $c\bar c$~-annihilation
amplitudes  to the bound state kinematics following
the so-called "on-energy-shell / off-mass-shell" prescription when
$\epsilon(p)\to \epsilon_{nS}=m_{nS}/2$ now remains independent of
the internal motion momentum $|\vec{p}|$ while being averaged with the
wave functions $\phi_{nS}(\vec{p})$ in (\ref{ave}).
%%%%%%%%%%%%%
This picture of the bound state dynamics underlies
the derivation of the relativistic Schr\"odinger-type wave equations
such as the quasipotential equation suggested by Todorov \cite{tod71} or
different variants thereof, {\it e.g.}, \cite{ge82}.
%%%%%%%%%%%%
In this prescription, the relativistic correction factor for
the electron-positron annihilation of vector charmonia is especially simple
because there are no additional momentum-dependent factors in the integrand,
while $\delta_{nS}^{J=0}(rel)$ is derived directly from the relation
\begin{eqnarray}
2[1+\delta_{nS}^{J=0}(rel)]^{1/2} = \frac{m_{c}^2}{\epsilon_{nS}^{2}}
\int \frac{d^{3}p}{(2\pi)^3}\phi_{nS}(p)
\frac{\epsilon_{nS}}{p}\log(\frac{\epsilon_{nS}+ p}{\epsilon_{nS}- p})
/\int\frac{d^{3}p}{(2\pi)^3}\phi_{nS}(p) \nonumber \\
= 2\frac{m_{c}^{2}}{\epsilon_{nS}}\int_{0}^{\infty}dr \psi_{nS}(r)
sin(\epsilon_{nS}r)/\psi_{nS}(0) \\
\simeq 2\frac{m_{c}^{2}}{\epsilon_{nS}^{2}}\cdot
[\psi_{nS}(0)-\frac{1}{\epsilon_{nS}^{2}}\psi_{nS}^{"}(0)+
\ldots]/\psi_{nS}(0),
\label{RiLe}
\end{eqnarray}
where masses $m_{nS}$ correspond to masses calculated without
spin-dependent corrections: $m_{1S}\simeq (3m_{J/\Psi}+m_{\eta_{c}})/4,
m_{2S}\simeq (3m_{\psi^{'}}+m_{\eta_{c}^{'}})/4$.

Keeping only the first term in asymptotic series for the Fourier
integral (\ref{RiLe}), we obtain, in the accord with the Riemann-Lebesgue
lemma, the simple approximate relation
\begin{eqnarray}
[1+\delta_{nS}^{J=0}(rel)]^{1/2} \simeq \frac{m_c^{2}}{\epsilon_{nS}^{2}}
\end{eqnarray}
demonstrating more strong dependence on the relativistic corrections of
the two-photon decay amplitude as compared to the leptonic one
\begin{eqnarray}
[1+\delta_{nS}^{J=1}(rel)]^{1/2} = 1- \frac{\epsilon_{nS} -m_c}{3 \epsilon_{nS}}.
\end{eqnarray}
Besides the relativistic corrections to be taken explicitly, we have observed
a significant contribution to our relation for ratios (\ref{double})
due to inclusion of the short-ranged spin-dependent interaction
which modifies the vector and pseudoscalar wave functions "at zero"
quite asymmetrically. However, for the linear combination (\ref{hfs})
of two ratios the "hyperfine" corrections are compensated up to terms
of the order ${\cal O}(\hat{V}^{2})$.

Collecting now all found corrections we get the resulting relation
between the widths of the lowest lying states of charmonia
\begin{eqnarray}
\frac{\Gamma_{\gamma \gamma}^{\eta_{c}^{'}}}{\Gamma_{\gamma \gamma}^{\eta_{c}}}
(\frac{m_{\eta_{c}^{'}}m_{2S}^{2}}{m_{\eta_{c}}m_{1S}^{2}})^2 +
3
\frac{\Gamma_{ee}^{\psi^{'}}}{\Gamma_{ee}^{J/\psi}}
(\frac{m_{\psi^{'}}m_{2S}(m_{1S}+m_{c})}{m_{J/\psi}m_{1S}(m_{2S}+m_{c})})^2
=4(\frac{\psi_{2S}(0)}{\psi_{1S}(0)})^2.
\label{main}
\end{eqnarray}
Hence, it follows that
\begin{eqnarray}
\frac{\Gamma_{\gamma \gamma}^{\eta_{c}^{'}}}{\Gamma_{\gamma \gamma}^{\eta_{c}}}
= 0.21 \pm 0.06,
\label{eqrat}
\end{eqnarray}
if we take $(\psi_{2S}(0)/\psi_{1S}(0))^2=0.653$ and $m_{c}=1.48$~GeV
according to \cite{bt81,eq95}, $\Gamma_{\gamma\gamma}^{\eta_{c}}$, masses
and leptonic widths from \cite{pdg02}.
A rather large uncertainty of the ratio obtained is largely due to experimental
errors of the measured leptonic widths.
%%%%%%%%%%%%%%%%%%%%%%%%%%%
It should be noted that unlike the very $\psi_{nS}(0)$'s
their calculated ratios are much less model-dependent .
In particular, the ratio entering into (\ref{main})
calculated with the Cornell ({\it i.e.}, the "linear+Coulomb")~type potential
is close to that calculated with the "running" $\alpha_{s}(r)$, which
makes the one-gluon-exchange potential softer at small distances, although
the very $\psi_{1S}(0)$'s differs about two times from each
other~\cite{eq95}.
We believe that mild smearing or regularization of the short-range quark
interactions providing formal finiteness of $\psi_{nS}(0)$'s following
from the relativistic equations will also leave their ratios relatively
intact.

As has been mentioned, the ratios of $\psi_{nS}(0)$ are dropped in (\ref{double})
and one can obtain a kind of the lower bound for the double width's ratio leaving
only one, beyond and next to unity, term in every infinite sum of the perturbation
corrections to $\Psi_{nS}^{J}(0)$ due to the contact spin-spin potential of
charmed quarks treated as a first order term of perturbation theory.
Including then already fixed relativistic corrections and assuming,
as earlier, the cancellation of the (static!) radiative corrections,
we obtain
\begin{eqnarray}
\frac{\Gamma_{\gamma \gamma}^{\eta_{c}^{'}}\Gamma_{ee}^{J/\psi}}
{\Gamma_{\gamma \gamma}^{\eta_{c}}\Gamma_{ee}^{\psi^{'}}}
(\frac{m_{\eta_{c}^{'}}m_{2S}m_{J/\psi}(m_{2S}+m_{c})}
{m_{\eta_{c}}m_{1S}m_{\psi^{'}}(m_{1S}+m_{c})})^2
\geq
\nonumber \\
(1+\frac{V_{11}^{J=0}+V_{22}^{J=0}}{m_{2S}-m_{1S}})^2/
(1+\frac{V_{11}^{J=1}+V_{22}^{J=1}}{m_{2S}-m_{1S}})^2 \simeq
1-8\frac{V_{11}^{J=1}+V_{22}^{J=1}}{m_{2S}-m_{1S}},
\end{eqnarray}
from where a new constraint follows
\begin{eqnarray}
\frac{\Gamma_{\gamma \gamma}^{\eta_{c}^{'}}}{\Gamma_{\gamma \gamma}^{\eta_{c}}}
\geq 0.1.
\label{ineqrat}
\end{eqnarray}
and where the numerical values~\cite{pdg02} for $m_{1S}^{J=1}$,
$m_{2S}^{J=1}$, defined earlier, and
$V_{11}^{J=1} = (m_{J/\psi}-m_{\eta_{c}})/4,~~
V_{22}^{J=1}=(m_{\psi^{'}}-m_{\eta_{c}^{'}})/4$,~~
were used.
The comparison of (\ref{eqrat}) and (\ref{ineqrat}) tells about a significant
role of sums
$<nS|{\hat V}^{J}|nS>$ over $n\geq 1,2$ in the definition
of the individual ratios of $\Gamma_{ee}^{J/\psi}/\Gamma_{ee}^{\psi^{'}}$
and especially of $\Gamma_{\gamma \gamma}^{\eta_{c}^{'}}/\Gamma_{\gamma \gamma}^{\eta_{c}}$.
%%%%%%%%%%%%%%%%%%%%%%%
%%%%%%%%%%%%%%%%%%%%%%%

To make contact with the available experimental data for radiative widths
of charmonia it is necessary to estimate also their relative branchings
referring to the studied hadronic decay channels.

The main assumptions underlying our estimations of the branching ratios
${\cal B}^{\eta^{'}_c}_{h}$~and ${\cal B}^{\eta_c}_{h}$~ entering into the
experimentally measured processes of the two-photon fusion producing
$\eta_c$ and $\eta_{c}^{'}$ and subsequent hadronic decays
$\eta_c~(\eta_{c}^{'})\to h$~, where $h\equiv K\bar{K}\pi$~\cite{cleo03}
or $h\equiv p\bar{p}$~\cite{e835}, are the following.
We assume a simple kinematic structure of the respective decay
amplitudes and an approximate dynamical assumption for the ratio of  relevant
couplings or, rather, complex form-factors in the considered vertices
\begin{eqnarray}
A(\eta_c(\eta_{c}^{'}) \to K\bar{K}\pi) = g(\eta_c(\eta_{c}^{'}) \to K\bar{K}\pi)\frac{1}{2}
\varphi_{K}^{\dag}\vec{\tau}\varphi_K \vec{\varphi}_{\pi},\\
A(\eta_c(\eta_{c}^{'}) \to p\bar{p}) = g(\eta_c(\eta_{c}^{'}) \to p\bar{p})
F_{c}(m_{c\bar{c}}^{2})\bar{u}(P_{p})\gamma_5 v(P_{\bar{p}}),
\label{prot}
\end{eqnarray}
\begin{eqnarray}
A(\eta_c(\eta_{c}^{'}) \to G_{m}G_{m}) = g(\eta_{c}(\eta_{c}^{'}) \to G_{m}G_{m})\varepsilon_{\mu\nu\rho\sigma}
e_{1}^{\mu}q_{1}^{\nu}e_{2}^{\rho}q_{2}^{\sigma},\\
\frac{|g(\eta_{c}\to h)|^2}{|g(\eta_{c} \to G_{m}G_{m})|^2} \simeq
\frac{|g(\eta_{c}^{'} \to h)|^2}{|g(\eta_{c}^{'} \to G_{m}G_{m})|^2},
\label{ampl}
\end{eqnarray}
where only isospin structure of the $ K\bar{K}\pi$ decay channel is indicated
in the first line
and the generalization to the $SU(3)$-symmetry can easily be written down.
We include the form-factor $F_{c}(Q^2)$ in (\ref{prot})
to mention about its possible variation depending on the time-like
momentum transferred in the interval
$m_{\eta_{c}}^{2} \leq Q^2 \leq m_{\eta_{c}^{'}}^{2}$.
We note that unlike the vector charmonia $J/\psi-$~ and $\psi^{'}$-decays,
the pseudoscalar decay
branchings ${\cal B}(\eta_{c}^{'}(\eta_{c}) \to (h,\gamma)(c\bar{c}))$
containing the charmed quarks in the final state, {\it e.g.},~ the decay
$\eta_{c}^{'}\to 2\pi \eta_{c}$,
are much less significant due to smallness of the coupling constant
$\alpha_{s}(m_{c}) \simeq 0.3$,~
as compared to the branching ratios
${\cal B}(\eta_{c}^{'}(\eta_{c}) \to h)$,~
 where $h$ denotes hadron states composed of the light $(u,d,s)$
quarks. Following a usual practice (or, alternatively,
the quark-hadron duality hypothesis for inclusive processes),
we identify the total width of these processes
with the width of the bound $c\bar{c}$-quark annihilation to pair of
"free" gluons $\Gamma(\eta_{c}(\eta_{c}^{'}) \to G_{m}G_{m})$.
Further, we attribute to gluons finite "effective" (or dynamical) mass
$m_{G}$ of the order 0.7~GeV, which was advocated in ref.~\cite{field02}
and in some earlier works cited therein, on the basis
of the detailed study of the experimental photon spectrum
in the inclusive reaction $J/\psi \to \gamma G_{m}G_{m} \to \gamma X$.
The evaluation of the important transition probabilities
$\eta_{c}(\eta_{c}^{'}) \to K\bar{K}\pi$, studied in several
experiments, see, {\it e.g.},~ \cite{cleo03} and further
references therein, has been performed with the help of
the integral relation for invariant 3-body phase space~\cite{dd02}
\begin{eqnarray}
\rho(m_{0} \to m_{1}+m_{2}+m_{3}) =\frac{1}{128\pi^3 m_{0}^{2}}
\int_{s_2}^{s_3}\frac{ds}{s}F(s,s_1,s_2,s_3,s_4),\nonumber \\
F(s,s_1,s_2,s_3,s_4) = [(s-s_1)(s-s_2)(s_3-s)(s_4-s)]^{0.5}
\end{eqnarray}
where $s_{2,1}=(m_1 \pm m_2)^2,~~ s_{4,3}=(m_0 \pm m_3)^2$ and we
choose $m_0=m_{\eta_{c}(\eta_{c}^{'})},~` m_1=m_2=m_{K},~~ m_3=m_{\pi}$.
After that, the integral which is the essential part of the ratio
${\cal B}_{K\bar{K}\pi}^{\eta_{c}^{'}}/{\cal B}_{K\bar{K}\pi}^{\eta_{c}}$
can easily be calculated numerically.

Including standard relativistic normalization factors of the initial states,
summing and averaging over spin degrees of freedom of the listed
amplitudes squared and using the assumed relation (\ref{ampl}),
we obtain
\begin{eqnarray}
\frac{{\cal B}_{h}^{\eta_{c}^{'}}}{{\cal B}_{h}^{\eta_{c}}}=
\begin{array}{ll}
0.83,~ \mbox{for}~h=K\bar{K}\pi, \\
0.65 \cdot(F_{c}(m_{\eta_{c}^{'}}^{2})/F_{c}(m_{\eta_{c}}^{2}))^2,~ \mbox{for}~ h=p\bar{p}.
\end{array}
\end{eqnarray}
Hence, our results are presented in Table 1, where
the upper bound for the $p\bar{p}$~-decay channel refers to the ratio
of the form-factors put equal to unity and
where we have included also some recent theoretical
predictions and experimental results for the widths of the $\eta_{c}(\eta_{c}^{'})$-to-$\gamma \gamma$
decays and their ratios.
\begin{table*}[htbp]
\setlength{\tabcolsep}{0.5cm}
\caption{\small Recent theoretical
and experimental results of the $\eta_c$ and $\eta_{c}^{'}$ two-photon
decay width~(the notations:~ $h\equiv \bar{K}K\pi$~
for ref.\cite{cleo03}~and $h\equiv \bar{p}p$~for ref.\cite{e760,e835}~,
the entry with superscript $(x)$~is our estimate using other works)}.
\label{tab1}
\begin{tabular*}{\textwidth}{@{}c@{\extracolsep{\fill}}cccc}
 \hline \hline
{\phantom{\Large{l}}}\raisebox{+.2cm}{\phantom{\Large{j}}}
&$\Gamma^{\eta_c}_{2\gamma}$~KeV &$\Gamma^{\eta^{'}_{c}}_{2\gamma}$~KeV&
$\Gamma^{\eta^{'}_c}_{2\gamma}/\Gamma^{\eta_c}_{2\gamma}$&
$\Gamma^{\eta^{'}_c}_{2\gamma}{\cal B}^{\eta^{'}_c}_{h}/
\Gamma^{\eta_c}_{2\gamma}{\cal B}^{\eta_c}_{h}$ \\
\hline\hline
{\phantom{\Large{l}}}\raisebox{+.2cm}{\phantom{\Large{j}}}
PDG \cite{pdg02}& { $7.5\pm 0.8$}&&&\\
{\phantom{\Large{l}}}\raisebox{+.2cm}{\phantom{\Large{j}}}
CLEO \cite{cleo00,cleo03}& { $7.6(0.8)(2.3)$}&&&0.18(0.05)(0.02)\\
{\phantom{\Large{l}}}\raisebox{+.2cm}{\phantom{\Large{j}}}
E760/E835 \cite{e760,e835}& { $6.7^{+2.4}_{-1.7}(2.3)$}&&&$\le$0.16\\
{\phantom{\Large{l}}}\raisebox{+.2cm}{\phantom{\Large{j}}}
L3 \cite{l3-99}& { $6.9(1.7)(2.1)$}&$\le$ 2&&\\
{\phantom{\Large{l}}}\raisebox{+.2cm}{\phantom{\Large{j}}}
DELPHI \cite{delphi98}&&&$\le$ 0.34&\\
\hline
{\phantom{\Large{l}}}\raisebox{+.2cm}{\phantom{\Large{j}}}
Gupta\cite{gupta96} ~&~10.94&&&\\
{\phantom{\Large{l}}}\raisebox{+.2cm}{\phantom{\Large{j}}}
M\"unz\cite{munz96} ~&~3.50$\pm$0.40&1.38$\pm$0.30&0.39$\pm0.10^{(x)}$&\\
{\phantom{\Large{l}}}\raisebox{+.2cm}{\phantom{\Large{j}}}
Chao \cite{chao97}~&~6-7 &2&$0.28-0.33^{(x)}$&\\
{\phantom{\Large{l}}}\raisebox{+.2cm}{\phantom{\Large{j}}}
Fabiano \cite{fab02}~&~8.18(0.57)(0.04)&&&\\
{\phantom{\Large{l}}}\raisebox{+.2cm}{\phantom{\Large{j}}}
Ebert \cite{efg03} ~&~5.5 &1.8&$0.33^{(x)}$&\\
{\phantom{\Large{l}}}\raisebox{+.2cm}{\phantom{\Large{j}}}
Kim\cite{kim04}~&~7.14$\pm$0.95&4.44$\pm$0.48&0.62$\pm0.10^{(x)}$&\\
{\phantom{\Large{l}}}\raisebox{+.2cm}{\phantom{\Large{j}}}
This work~&&1.6$\pm0.5^{(x)}$&0.21$\pm$0.06&0.18~$,h\equiv K\bar{K}\pi$ \\
{\phantom{\Large{l}}}\raisebox{+.2cm}{\phantom{\Large{j}}}
          &&&&$ \leq 0.15,~h\equiv p\bar{p}$\\
 \hline\hline
\end{tabular*}
\end{table*}

The behaviour of the unitary-singlet, pseudoscalar form-factors in
the hadronic transition vertices is of considerable interest for
the understanding of mechanisms of sequential processes
$\eta_{c}(\eta_{c}^{'}) \to G_{m}G_{m} \to light~ hadrons$.
The closeness of our estimated $R(\eta_{c}^{'}/\eta_{c})$~
decay-ratio in the $K\bar{K}\pi$ channel, with no additional form-factors
included, to the CLEO data looks
intriguing and needs more investigation to be understood. At any rate,
in the $p\bar{p}$ channel one should expect a stronger
dependence of the result on the $(m_{\eta_{c}}^2/m_{\eta_{c}^{'}}^2)^n$~ratio,
where the effective power $n$ is expected to be two units larger in the
decay amplitude for two-baryon final state as compared to the transitions
into the two-meson states, according the quark-counting rules.
The two-gluon state mediating the $c\bar{c}$~- and
$q^{n}\bar{q}^n$~- states, where $q=u,d,s$,
would provide a new testing ground for checking
the generalized parton approach or the diquark model which were successful
in the description of the two-photon annihilation processes, like
$\gamma \gamma \to p\bar{p}$, etc. Therefore, a further study of
the reactions $p\bar{p} \to \eta_{c}(\eta_{c}^{'}) \to \gamma \gamma$
(or $K\bar{K}\pi$) with better statistics and accuracy is of
interest, {\it e.g.}, via the ongoing experiments at FNAL or
at a planned antiproton storage ring at GSI.

{\bf 3.}~~Finally, we note that relation (\ref{main}) can be applied to
any pairs of the "hyperfine-split" radial-excited states.In particular,
using it for pairs of ratios $R_{3S/1S}$ and $R_{3S/2S}$ with the needed
input values $m_{\psi^{"}}=4039$~MeV, $\Gamma_{ee}^{\psi^{"}}=.89\pm.08$~keV~\cite{seth04,bes02},
$|\psi_{3S}(0)/\psi_{1S(2S)}(0)|^2 \\
=.56(.86)$~\cite{eq95}
we obtain the estimation $m_{\eta_{c}^{"}}\simeq 4003$~MeV and
$\Gamma_{\gamma \gamma}^{\eta_{c}^{"}}/\Gamma_{\gamma \gamma}^{\eta_{c}}
\simeq .22$~of the "naked"~({\it i.e.}, without possible hadronic corrections
due to virtual, open-charm intermediate $(D\bar{D^{*}}+\bar{D}D^{*})$~- states)
parameters of the still to be observed $\eta_{c}^{"}$-resonance.

Our main results, relation~(\ref{main}) and the numerical entries in
Table 1, demonstrate a considerable suppressing effect of the relativistic
and "hyperfine" spin-dependent corrections on the recently observed two-photon decay of
the $\eta_{c}^{'}(3640)$~-resonance. If it is true, this effect should
also display itself in the total width $\Gamma_{tot}^{\eta_{c}^{'}}$
represented by the decay into two gluons, either massless or effectively
massive. Using the average value $\Gamma_{tot}^{\eta_{c}}=
32.3 \pm 2.2$~MeV of the CLEO~\cite{cleo03} and BaBar~\cite{babar03} results,
we have got the estimate of the total width of the $\eta_{c}^{'}(3640)$- resonance
\begin{eqnarray}
\Gamma_{tot}^{\eta_{c}^{'}}=
\begin{array}{ll}
(32.3\pm 2.2)\cdot(.21\pm.06)\simeq 6.8\pm 2.0~MeV,~ \mbox{for}~m_{G}=0, \\
(32.3\pm 2.2)\cdot(.21\pm.06)\cdot((1-4m_{G}^{2}/m_{\eta_{c}^{'}}^2)/
(1-4m_{G}^{2}/m_{\eta_{c}}^{2}))^{3/2} \nonumber \\
\simeq 7.8\pm 2.3~MeV,~ \mbox{for}~m_{G}\simeq .7~GeV .
\end{array}
\end{eqnarray}
A more precise measurement of this important parameter, as compared to
the published~\cite{cleo03} result
$\Gamma_{tot}^{\eta_{c}^{'}}=6.3_{-4.0}^{+12.4}$~MeV, would be very desirable.

\section*{Acknowledgements}
S.G. thanks Prof.J.Rembielinski
and the staff of the Department of Theoretical Physics
for warm hospitality and partial support during his visits to
University of Lodz.

Partial support of this work by the Bogoliubov-Infeld foundation is
gratefully acknowledged.

\end{document}